\documentclass[aps,prl,twocolumn,groupedaddress]{revtex4-1}

\usepackage{graphicx}
\usepackage{amsfonts}
\usepackage{bm}
\usepackage{amsmath}
\usepackage{amssymb}
\usepackage{verbatim}
\usepackage{comment}

\usepackage[]{morefloats}

\newcommand{\vk}{{\mbox{\boldmath$k$}}}

\newcommand{\vv}{{\mbox{\boldmath$v$}}}

\newcommand{\vrx}{{\mbox{\boldmath$r$}}}
\newcommand{\vvF}{{\mbox{\boldmath$v$}}}

\newcommand{\vsigma}{\mbox{\boldmath$\sigma$}}

\newcommand{\vQ}{{\mbox{\boldmath$Q$}}}

\begin{document}

\title{Current inversion at the edges of a chiral $\bm{p}$-wave superconductor}


\author{Adrien Bouhon}
\email[]{abouhon@phys.ethz.ch}
\author{Manfred Sigrist}
\affiliation{Institute for theoretical physics, ETH Zurich}


\date{\today}

\begin{abstract}
Motivated by Sr$_2$RuO$_4$, edge quasiparticle states are analyzed based on the self-consistent solution of the Bogolyubov-de Gennes equations for a topological chiral $p$-wave superconductor. Using a tight-binding model of a square lattice for the dominant $\gamma$-band we explore the non-trivial geometry and band structure dependence of the edge states and currents. As a peculiar finding we show that for high band fillings currents flow in reversed direction comparing straight and zigzag edges. We give a simple explanation in terms of the positions of the zero-energy bound states using a quasi-classical picture. We also show that a Ginzburg-Landau approach can reproduce these results. Moreover, the band filling dependence of the most stable domain wall structure is discussed. 
\end{abstract}




\maketitle

Since the discovery of superconductivity in Sr$_2$RuO$_4$ numerous experiments revealed the unconventional nature of the pairing state, many suggesting the spin-triplet chiral $p$-wave (CPW) state as the strongest candidate \cite{Maeno-Mackenzie, Phys-Today,JPSJ-Maeno}. This state belongs to the two-dimensional irreducible representation $ E_u $ of the tetragonal point group ($D_{4h}$) with a gap function parametrized by the vector $\bm{d}(\vk) = -\mathrm{tr}\{ \hat{\Delta}_{\vk} i \hat{\sigma}_y \hat{\bm{\sigma}}\}/2 =  \hat{z} \Delta_0 \left( k_x\pm i k_y\right)/k_F   $, where $\hat{\Delta}_{\vk}$ is the gap matrix in spin space and $\Delta_0$ is the gap magnitude \cite{BalianWerthamer,Leggett1975}. 
It was pointed out that chiral superconducting phases are topological superconducting phases which can be characterized through the topological invariant associated to their ground states \cite{Balatskii1986,Volovik1992,Volovik1997,Volovik,MatSig2001}. 
Since time reversal symmetry is broken the topology is distinguished by the Chern number  ($C_1 \in \mathbb{Z}$). 

An important consequence of the topology is that spatial defects (sample surfaces, domain walls etc), where the Chern number changes, can host quasiparticle bound states whose energy eigenvalues cross the energy gap and connect the two separate sectors of the bulk spectrum. 
Generally in a chiral phase such a gap crossing 
yields an imbalance between certain momentum directions along the defect such that the bound states can generate a quasiparticle flow and local currents. 
In particular, in a CPW state we expect such surface currents generating local magnetic fields. 
While edge states have been observed by tunnelling spectroscopy in Sr$_2$RuO$_4$ \cite{Kashiwaya2011}, attempts to observe chiral edge currents have failed so far \cite{Kirtley,Budakian,Curran2014}. 
Our aim is to shed light on this question based on a microscopic study of the edge states in the CPW state. 

Using a tight-binding model of a square lattice for the $\gamma$-band of Sr$_2$RuO$_4$ \cite{Thomale2013} and, assuming the CPW state, we solve the Bogolyubov-de Gennes equations self-consistently. We study the geometry and band structure dependence of the edge states and discuss the current pattern that they generate. It turns out that edge states and currents can strongly depend on the band filling and the orientation of the surfaces. For certain conditions the edge state spectrum qualitatively changes and reverses the edge current directions and can give rise to an unusual current pattern. This modification can also be straightforwardly reproduced within a Ginzburg-Landau formulation. 
The effective tight-binding mean field Hamiltonian reads,
\begin{equation}
\label{Ham1}
	\mathcal{H} = \sum\limits_{ij} \epsilon_{ij} c^{\dagger}_{i} c_{j} + \Delta_{ij} c^{\dagger}_{i} c^{\dagger}_{j} + \Delta^*_{ji} c_{i} c_{j} \;,
\end{equation}
where $\epsilon_{ij} = \left(\epsilon_0 - \mu \right) \delta_{ij} - t_{\gamma} \delta_{\langle ij \rangle} - t'_{\gamma} \delta_{\langle\langle ij \rangle\rangle}$ is the tight-binding dispersion relation of the $\gamma$-band with onsite ($\epsilon_0 - \mu$), nearest-neighbor ($t_{\gamma}$), and next-to-nearest-neighbor ($t'_{\gamma}$) hopping terms. The mean field gap function is defined through the gap equation $\Delta_{ij} = - g_{ij} \langle c_{i} c_{j} \rangle $, with the pairing $g_{ij} = g^p \delta_{\langle ij \rangle} $, where we restrict to nearest-neighbor (intersublattice) pairing interactions. 
The homogeneous system follows the Hamiltonian, 
\begin{equation}
\mathcal{H}^{\mathrm{bulk}} = \sum_{\vk} \bm{c}^{\dagger}_{\vk} h_{\vk}  \bm{c}_{\vk}, \quad h_{\vk}=\left[\begin{array}{cc} \epsilon_{\vk} & \Delta_{\vk}\\ \Delta^*_{\vk} & - \epsilon_{\vk} \end{array}\right]
\label{Ham-bulk}
\end{equation}
 with $\bm{c}_{\vk} = (c_{\vk} ~ c^{\dagger}_{-\vk}  )^T$. The dispersion relation for the $\gamma$-band is now given by $ \epsilon_{\vk} = \epsilon_{0} -\mu -2 t_{\gamma} (\cos k_x a + \cos k_y a) - 4 t'_{\gamma} \cos k_x a \cdot \cos k_y a$, and the gap function is $\Delta(\vk) =  \Delta_x \sin k_x a+ \Delta_y\sin  k_y a  $, with $(\Delta_x,\Delta_y) = \Delta_0 (1,\pm i) $ ($ \Delta_0 \in \mathbb{R}$). 

%
%

In the following we solve the Bogolyubov-de Gennes equation for a ribbon shaped system based on the Hamiltonian (\ref{Ham1})
for two basic orientations: one parallel to a principal crystal direction having straight edges, and the other along diagonal direction, with zigzag edges. We assume translational invariance and impose periodic boundary conditions along the ribbon direction,
introducing the Bloch wave vector parallel to the edge, $k_{\parallel}$, to label the eigenstates and energy eigenvalues. For a ribbon with $N$ sites in the perpendicular direction we then have $N$ equations,
\begin{equation}
	\sum\limits_{j} \left[\begin{array}{cc} \epsilon_{ij}(k_{\parallel}) & \Delta_{ij}(k_{\parallel}) \\ \Delta^*_{ji}(k_{\parallel}) & -\epsilon_{ij}(-k_{\parallel}) \end{array}\right] \left[\begin{array}{c}	 u^n_{j,k_{\parallel}} \\ v^n_{j,k_{\parallel}} \end{array}\right] = E_{n,k_{\parallel}} \left[\begin{array}{c}	 u^n_{i,k_{\parallel}} \\ v^n_{i,k_{\parallel}} \end{array}\right] \;, 
\end{equation}
which lead to a spectrum of $2N$ eigenvalues $\left\{E_{n,k_{\parallel}}\right\}_{n=1,\dots,2N}$ for every $k_{\parallel}$ belonging to the reduced Brillouin zone $(-\Lambda_{\parallel},\Lambda_{\parallel}]$. The local spectral function is then given by,
\begin{equation}
	A(E,k_{\parallel},\vrx_i) = \sum_{n=1}^{2N} \left(\vert u^n_{i,k_{\parallel}}\vert^2 + \vert v^n_{i,k_{\parallel}}\vert^2\right) \delta(E -E_{n,k_{\parallel}}) \;.
\end{equation}
%

%
%



%


%
\begin{figure}[t!]
\centering
\begin{tabular}{c}
	\includegraphics[width=0.33\textwidth]{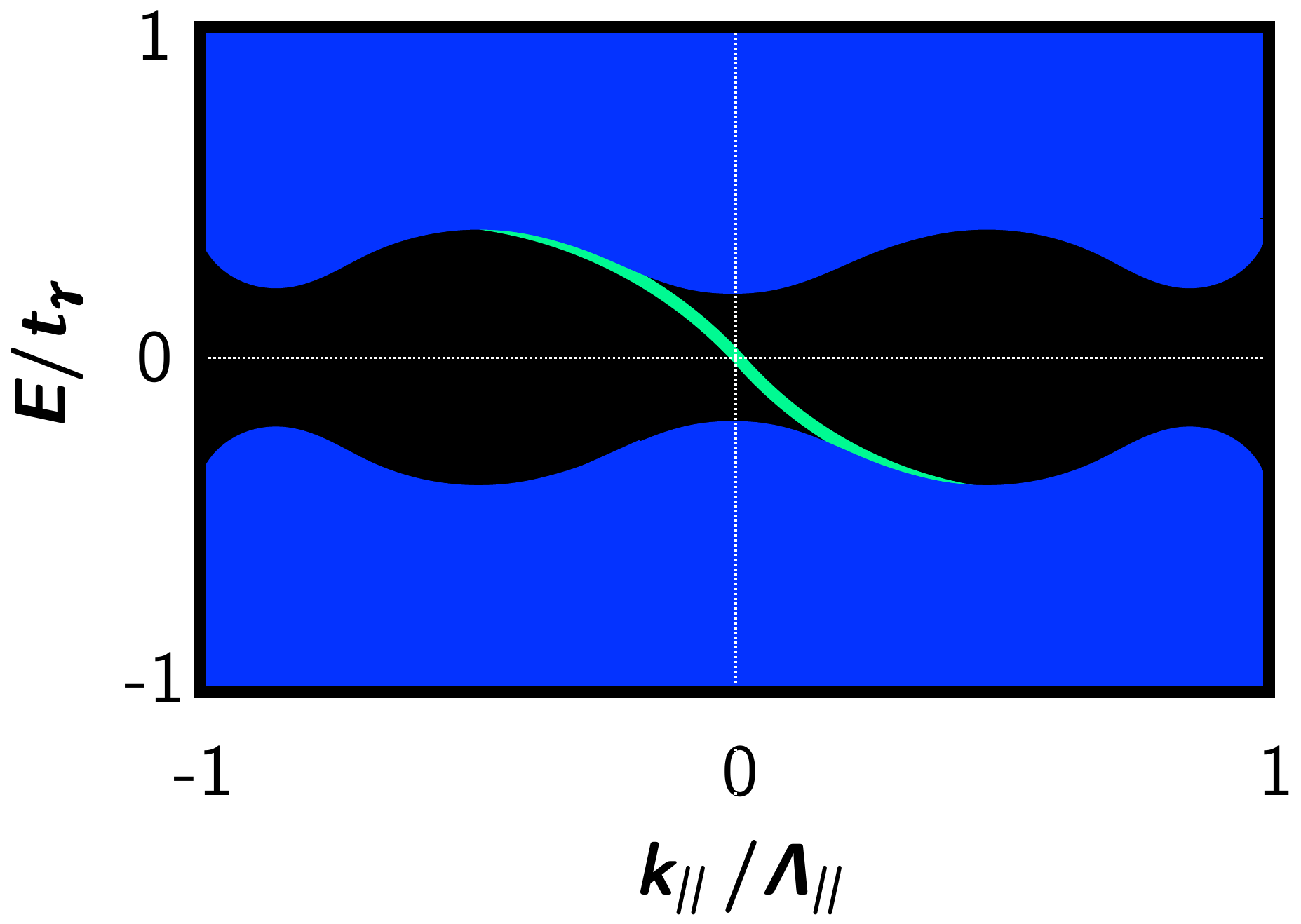} \\
	(a) $\theta_{\mathrm{Edge}}=0^{\circ}$. \\
	\includegraphics[width=0.33\textwidth]{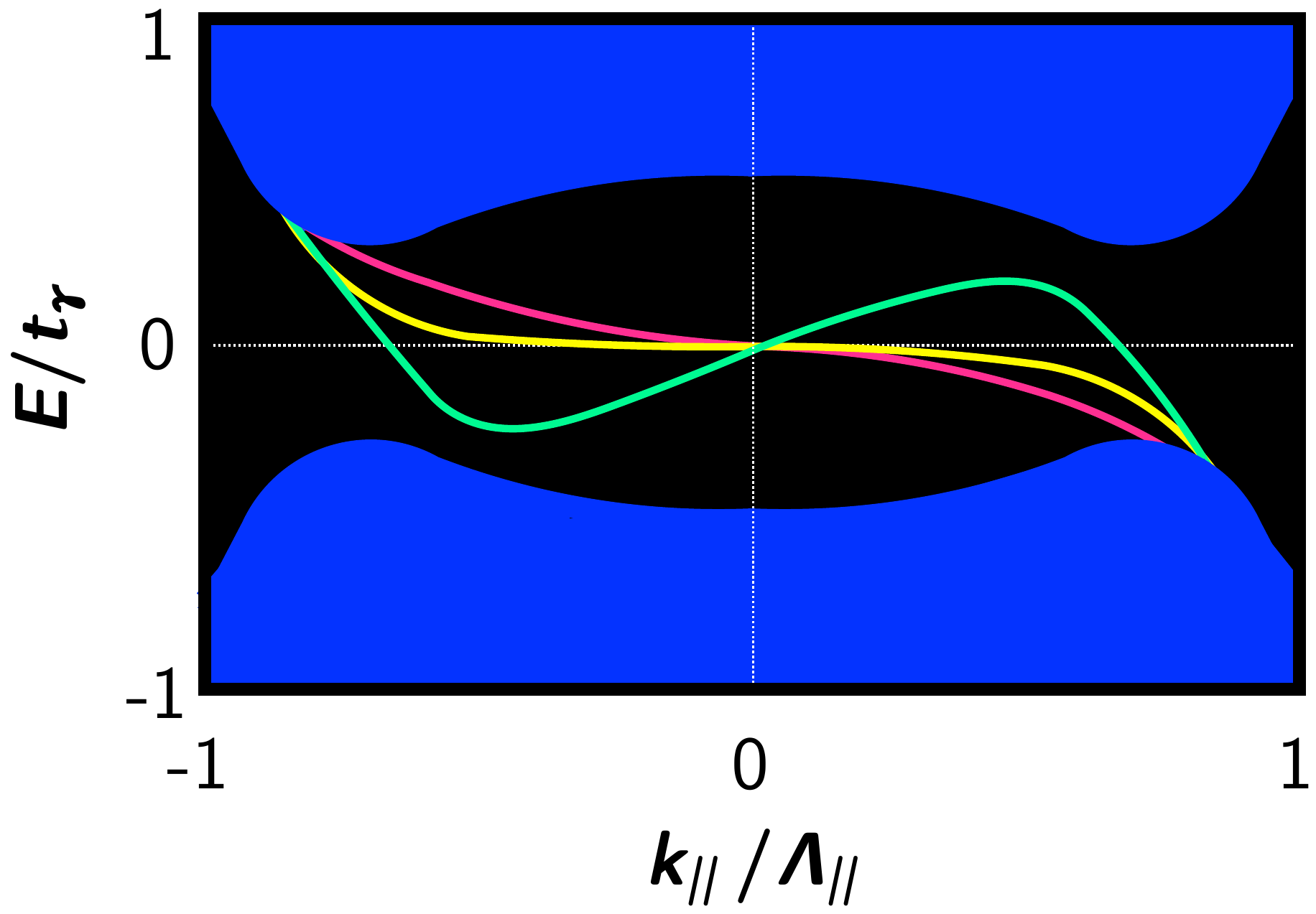}\\
	(b)  $\theta_{\mathrm{Edge}}=45^{\circ}$.
\end{tabular}
\caption{\label{D_ES} Spectral function (drawing made from the numerical self-consistent solution of the BdG equation) at an edge of a ribbon as a function of the Block wave vector parallel to the edge, $k_{\parallel} \in (-\Lambda_{\parallel},\Lambda_{\parallel}]$, and the quasiparticle energy, $E$. The coordinates are chosen such that the normal vector at the surface $\bm{n}=(1,0,0)$ is pointing outwards. (a) Straight edge, i.e. the angle of the edge with respect to the main crystal direction is $\theta_{\mathrm{Edge}}=0^{\circ}$. (b) zigzag edge, i.e. with  $\theta_{\mathrm{Edge}}=45^{\circ}$, at different fillings : low filling (red line), high filling (green line), and transition filling (yellow line).} 
\end{figure}
\begin{figure}[t!]
\begin{tabular}{c} 
\includegraphics[width=0.23\textwidth]{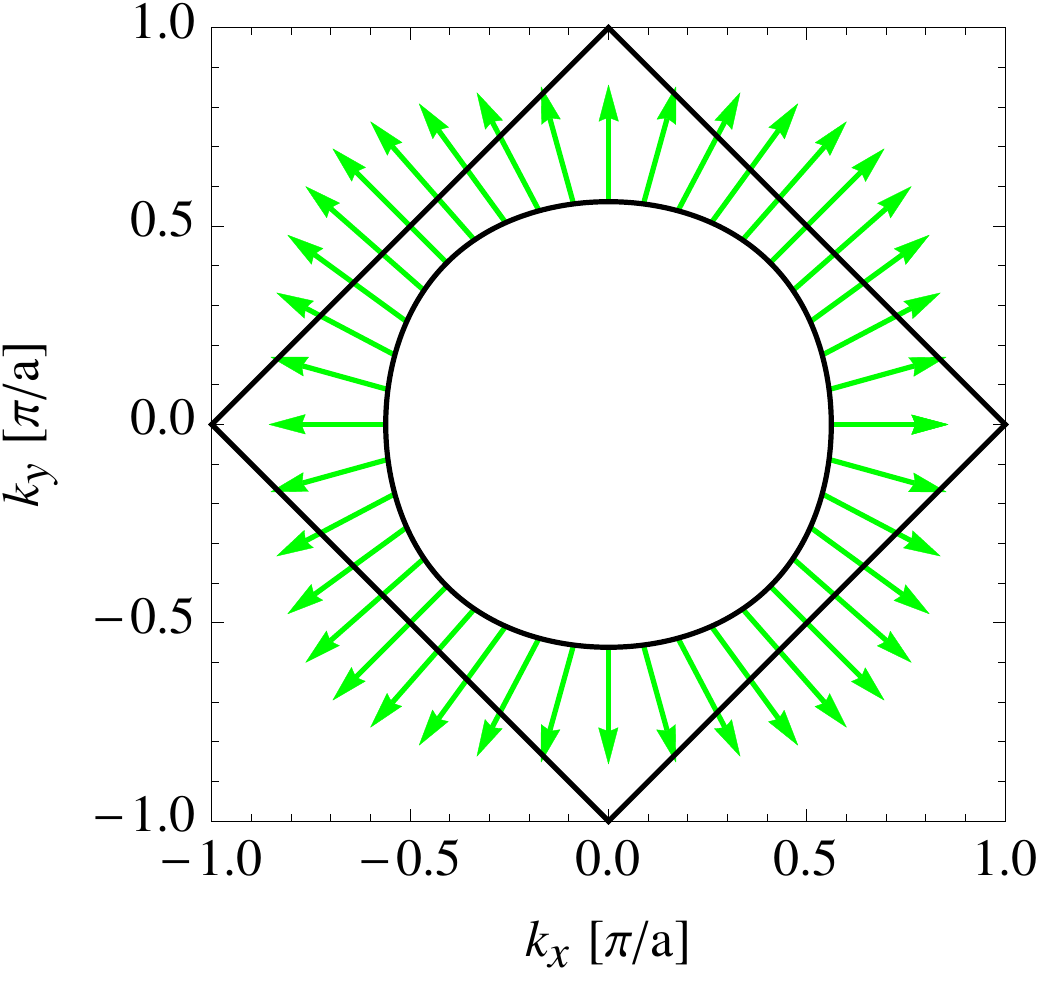} 
\includegraphics[width=0.23\textwidth]{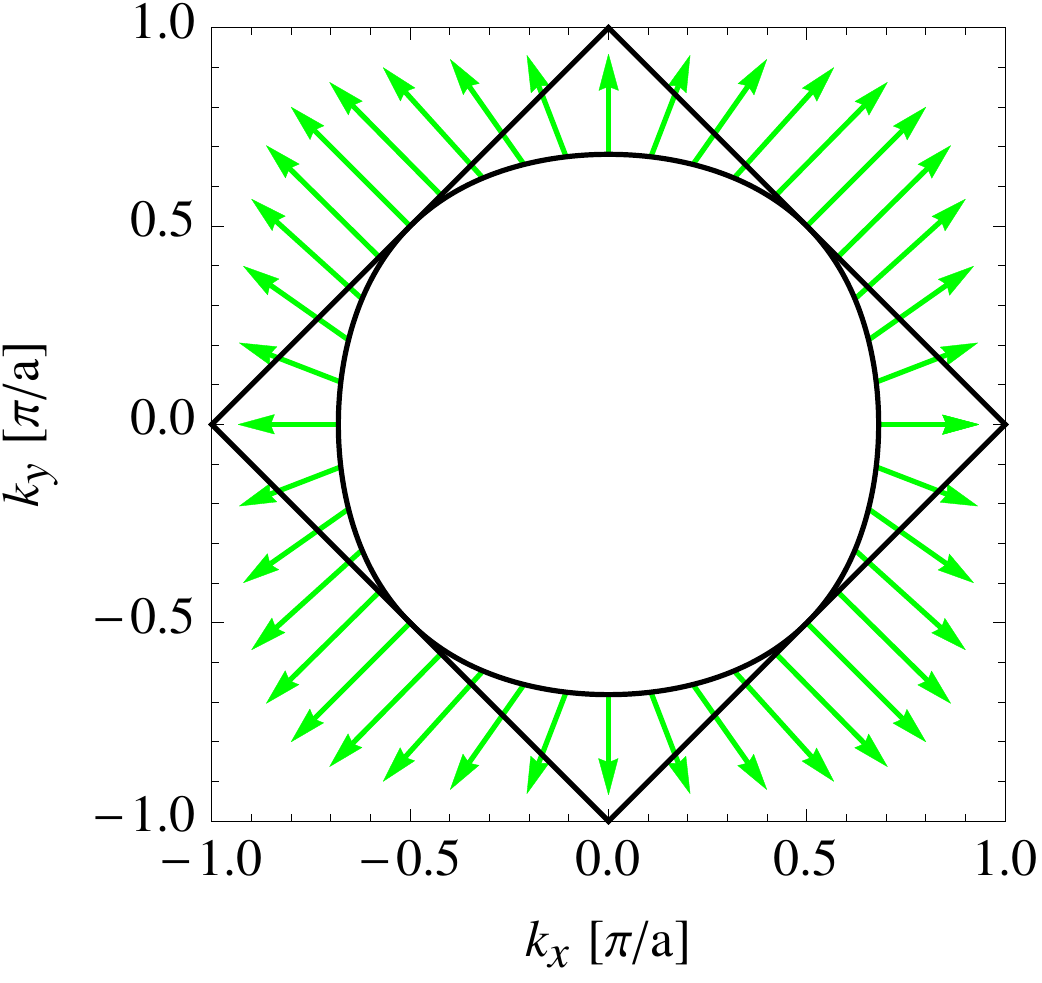}\\
(a) $\mu-\epsilon_0  < 0$.  \hspace{2cm} (b) $\mu-\epsilon_0  = 0$. \\
\includegraphics[width=0.235\textwidth]{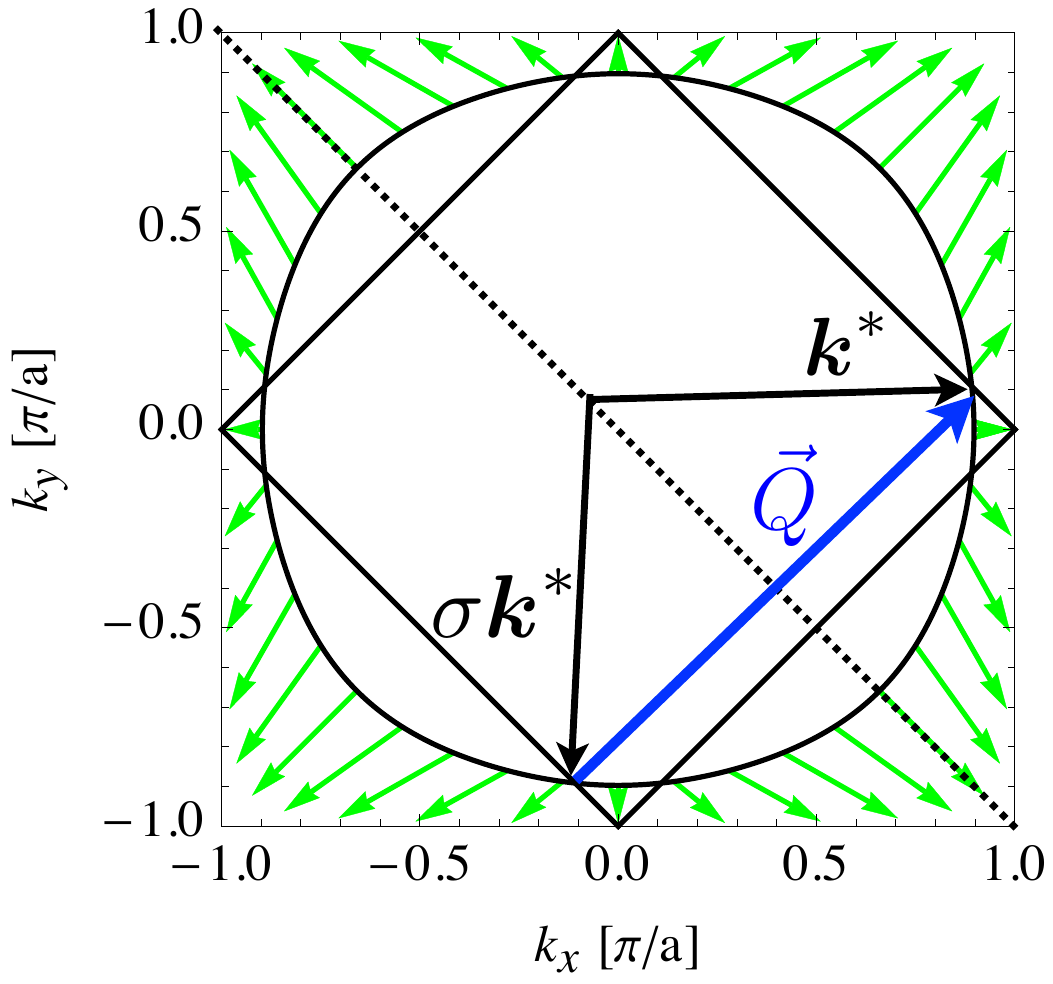} \\
(c) $\mu-\epsilon_0  > 0$. 
\end{tabular}
\caption{\label{FS} CPW gap function of the $\gamma$-band at the Fermi level, $\Delta(\vk_F)/\Delta_0 =\sin k_{F,x} a + i \sin k_{F,y} a$, represented as a vector in the complex plane $(k_x,ik_y)\in \mathbb{C}$, for different fillings. (a) Low filling: there is no crossing point between the Fermi surface and the nesting diamond. (b) Transition filling: the Fermi surface touches the nesting diamond. (c) High filling: there are four pairs of crossing points between the Fermi surface and the nesting diamond, which we write $\vk^*$. We note that in addition to the $\pi$-phase shift of the gap function under inversion, at high filling there is a $\pi$-phase shift of the gap function between each two crossing points $\vk^*$ and $\sigma \vk^*$, where $\sigma$ is the reflection operation for a mirror plane along the diagonal of the Brillouin zone. They are connected through $\vk^*=\sigma \vk^*+\vQ$, with the "nesting" vector $\vQ=\left(\pi/a,\pi/a \right)$. 
}
\end{figure}
%

The clearest signature of the topological nature of the superconducting state can be seen in the
edge states, as displayed in Fig. \ref{D_ES}  through the local spectral function at one edge of the ribbon. Panel (a) shows the situation for the straight edge ($\theta_{\rm Edge} = 0^{\circ}$) with one spin degenerate subgap edge mode crossing from the upper to the lower continuum. The qualitative behavior does not depend on band filling. This is different for the case of the zigzag edge ($\theta_{\rm Edge} = 45^{\circ}$) where an intriguing filling dependence can be observed in the panel (b). The three subgap spectra correspond to the fillings 
depicted in Fig.~\ref{FS}. For small filling the Fermi surface lies within the ''nesting'' diamond (Fig.~\ref{FS}(a)) and the CPW state generates a spectrum analogous to the case of panel (a) with one zero-crossing of the edge state (red curve). On the other hand, if the band 
crosses the diamond  (Fig.~\ref{FS}(c)) then we find three momenta $ k_{\parallel} $ with zero-energy states (green curve). The transition between the two limits is continuous where the subgap state energy dispersion is flat at the crossing point $ k_{\parallel} = 0 $ (yellow curve), coinciding with the band filling for which the Fermi surface just touches the diamond (Fig.~\ref{FS}(b)). 

The unexpected behavior for the large band filling (green curve in Fig.\ref{D_ES}) is caused by the phase structure of the gap function on the Fermi surface plotted in Fig.~\ref{FS}(c). It is straightforward to derive from a quasi-classical approach that a phase shift of $\pi $ of the gap function between the momenta incident to the edge and scattered from the edge, $ \vk $ and $ \vk'= \sigma \vk $ with $ \sigma $ being the reflection operation using a mirror plane parallel to the sample edge (for specular scattering), leads to the existence of a zero-energy state with the momentum $ \vk_{\parallel} = (\vk + \vk')/2 $. 


The gap function based on nearest-neighbor pairing acquires a $ \pi $-phase shift for straight edges, generally, and zigzag edges at small filling (Fig.~\ref{FS}(a)) only, if $ k_{\parallel} = 0 $. The crossing points of the Fermi surface on the nesting diamond in Fig.~\ref{FS}(c) play a special role for the zigzag edge, because they correspond to momenta with $  \vk' - \vk = \vQ = (\pi/a, \pi/a ) $ where we find for inter-sublattice pairing $ \Delta (\vk') = \Delta(\vk+\vQ) =  - \Delta (\vk) $, i.e. for these momenta also a $ \pi $-phase difference appears leading to additional zero-energy states. 

It is important to notice that this modification of the spectrum is not a topological feature. The topology of the superconducting phase is described by the Chern number given by the integral
\begin{eqnarray}
C_1= \dfrac{1}{4\pi} \int_{T^2} d^2k~ \hat{m}_{\vk} \cdot \left( \partial_{k_x}\hat{m}_{\vk} \times \partial_{k_y} \hat{m}_{\vk}\right)   \;.
\end{eqnarray}
where the bulk Hamiltonian (\ref{Ham-bulk}) is represented as $h_{\vk} = \bm{m}_{\vk} \cdot \hat{\vsigma}$, from which we obtain the unit vector $\hat{m}_{\vk} = \bm{m}_{\vk}/\vert \bm{m}_{\vk}\vert$. The difference of the Chern numbers on the two sides of the edge, i.e. sample and vacuum, coincides to the number of states in
the spectrum traversing from the upper to the lower bulk continuum when scanning the $ k_{\parallel} $ from $ - \Lambda_{\parallel} $ to $ + \Lambda_{\parallel} $. 
In our case $ \Delta C_1 = C_1({\rm outside}) - C_1({\rm inside}) = 0 -1= -1 $ implies that one state moves down as can be seen in Fig.\ref{D_ES} (a,b). Indeed this feature remains unchanged even in the case of three zero-crossings.


We add here a remarkable observation concerning the current density at different edges. To motivate this we analyze qualitatively the contribution of the subgap bound states to the edge current density which we express as $ J_{\parallel} \propto \sum_{k_{\parallel}} n_{k_{\parallel}} v_{k_{\parallel}} $. Note that the velocity $ v_{k_{\parallel}}$ is determined by the quasiparticle dispersion at Fermi surface and $ v_{-k_{\parallel}} = - v_{k_{\parallel}} $. For $ T=0 $ only states with negative energy have $ n_{k_{\parallel}} \neq 0 $. For straight edges these are only bound states with $ k_{\parallel} > 0 $ for all band fillings, as is obvious from Fig.\ref{D_ES}(a). On the other hand, for zigzag edges we see a change of the momentum distribution. While for small filling the situation is identical to the straight edge, for large filling the distribution of the occupied edge states shifts to negative $ k_{\parallel} $ (Fig.\ref{D_ES}(b)). From this simple discussion we anticipate a reversal of the current density for zigzag edges at high filling, since the Fermi surface topology is unchanged (Fig.\ref{FS})  and $ v_{k_{\parallel}} $ keeps its momentum dependence qualitatively over the range of band filling considered here. 
This qualitative observation is indeed confirmed by our detailed numerical analysis including all contributions the edge currents.  By interpolation we can state that the edge current density has to vanish at an intermediate orientation, i.e. for $0 < \theta_{\mathrm{Edge}} < 45^{\circ}$, in this case. 

\begin{figure}[t!]
\centering
\begin{tabular}{cc}
	\includegraphics[width=0.19\textwidth]{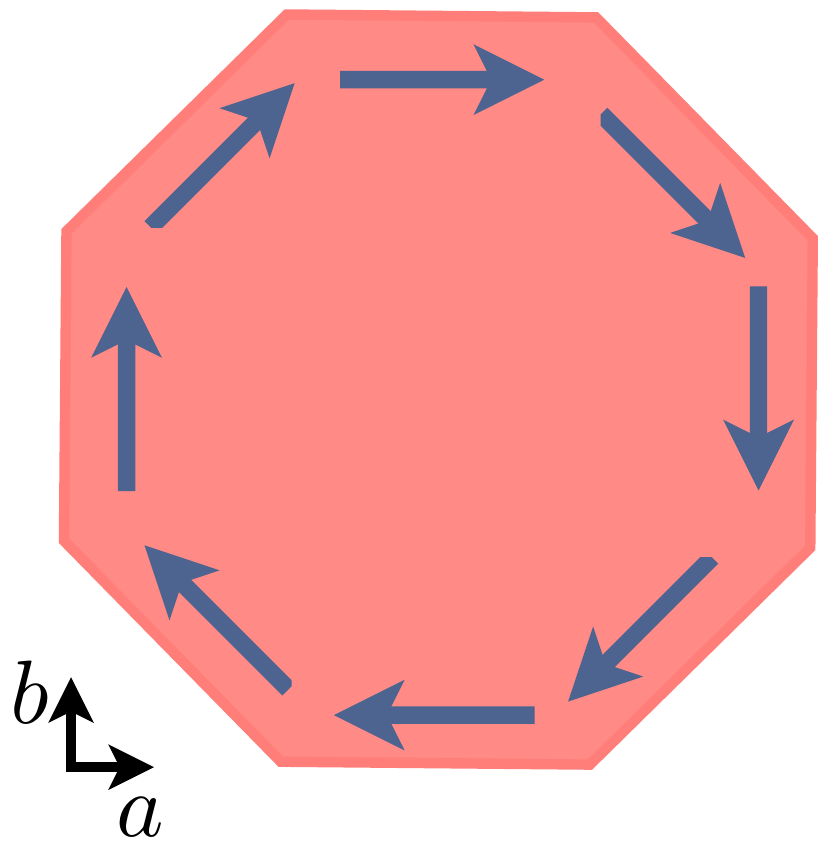} $\quad$&$\quad$
	\includegraphics[width=0.19\textwidth]{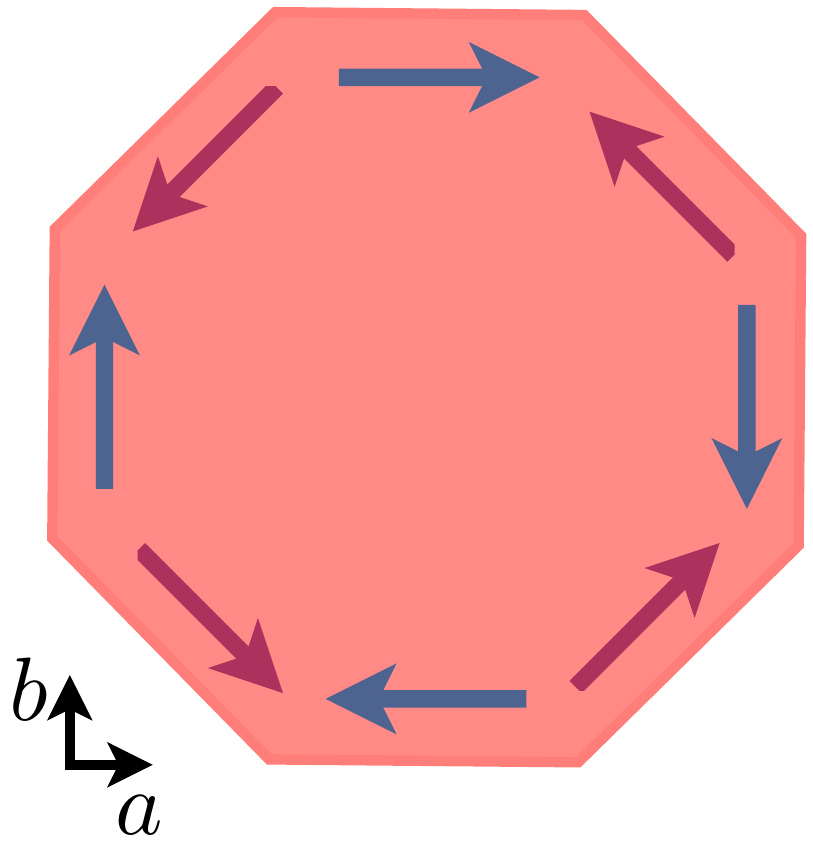}\\
	(a) Low filling. & (b) High filling.
\end{tabular}
\caption{\label{EC_Low_High} Schematic edge current pattern for an octagonally shaped sample at low (a) and high filling (b). If it is a CPW superconductor Sr$_2$RuO$_4$ belongs to the regime (b).} 
\end{figure}
It is illustrative to plot schematically the edge current patterns in a octagonally shaped sample corresponding to the low and high filling regimes in Fig. \ref{EC_Low_High} (a) and (b) respectively. While at low filling the edge current has a fully connected current circuit analogue to the quantum Hall state, at high filling the edge current now alternates between the straight and zigzag edges. In the latter case the overall current pattern would be somewhat more complex. Note, however, that edge currents are compensated over a distance of London penetration depth so as to screen the induced magnetic fields. 
The filling for Sr$_2$RuO$_4$ is rather large$-$a typical estimate for the $\gamma$-band is $ (\mu- \epsilon_0)/t_{\gamma} \approx 1.4$ \cite{Bergemann}. Thus, if it realizes the CPW state we expect this material to exhibit most likely the edge current pattern of Fig. \ref{EC_Low_High} (b). 

 We now turn to the Ginzburg-Landau (GL) formulation which also accounts for edge currents. The GL free energy functional can be expanded in  the two complex order parameter components $ (\Delta_x,\Delta_y)$ belonging to the irreducible representation $E_u$ of the point group $ D_{4h} $. We focus here on the gradient terms of the free energy density, needed to express the current density,
\begin{equation}
\label{eq_GL} 
f_{\rm grad} =  \sum_{\mu,\nu=x,y} [ K_{\mu \nu}  |\Pi_{\mu} \Delta_{\nu}|^2 + \tilde{K}_{\mu \nu} \{ (\Pi_\mu \Delta_\nu)^* (\Pi_{\bar{\nu}} \Delta_{\bar{\mu}}) +c.c. \}]
 \end{equation}
with $ \Pi_{\mu} = \frac{\hbar}{i} \partial_{\mu} + \gamma A_{\mu} $ ($ \gamma = 2e/c $), $ A_{\mu} $ the vector potential and
$ (\bar{x},\bar{y})  = (y,x)  $ \cite{Sigrist1991}. Within a weak-coupling approach for our tight-binding model we derive the coefficients $ K_{\mu \nu} $ and
$ \tilde{K}_{\mu \nu} $. These are given by the following averages over the Fermi surface $ K_{\mu \nu} = K_0 \langle N_0 v_{F\mu}^2 \phi_{\nu}^2 \rangle_{FS} $ and
$ \tilde{K}_{\mu \nu} = K_0 \langle N_0 v_{F \mu} v_{F \bar{\nu}} \phi_{\nu} \phi_{\bar{\mu}}   \rangle_{FS} $ with $ K_0 $ a common constant. The density of states is
given by $ N_0(\vk) \propto |\vk_F| / |\vvF_F| $, the Fermi velocities are defined as $ \vv_F = \left. \nabla_{k} \epsilon(\vk) \right|_{\vk=\vk_F} $, and the gap lattice form factors are given by $\phi_{\nu} = \sin k_{\nu} a$ for nearest-neighbor pairing. Note that by symmetry, $ K_{xx} = K_{yy} $, $ K_{xy}=K_{yx} $, $ \tilde{K}_{xx}=\tilde{K}_{yy}=\tilde{K}_{xy}=\tilde{K}_{yx} $
and all coefficients are positive. 

\begin{figure}[t]
\includegraphics[width=0.3\textwidth]{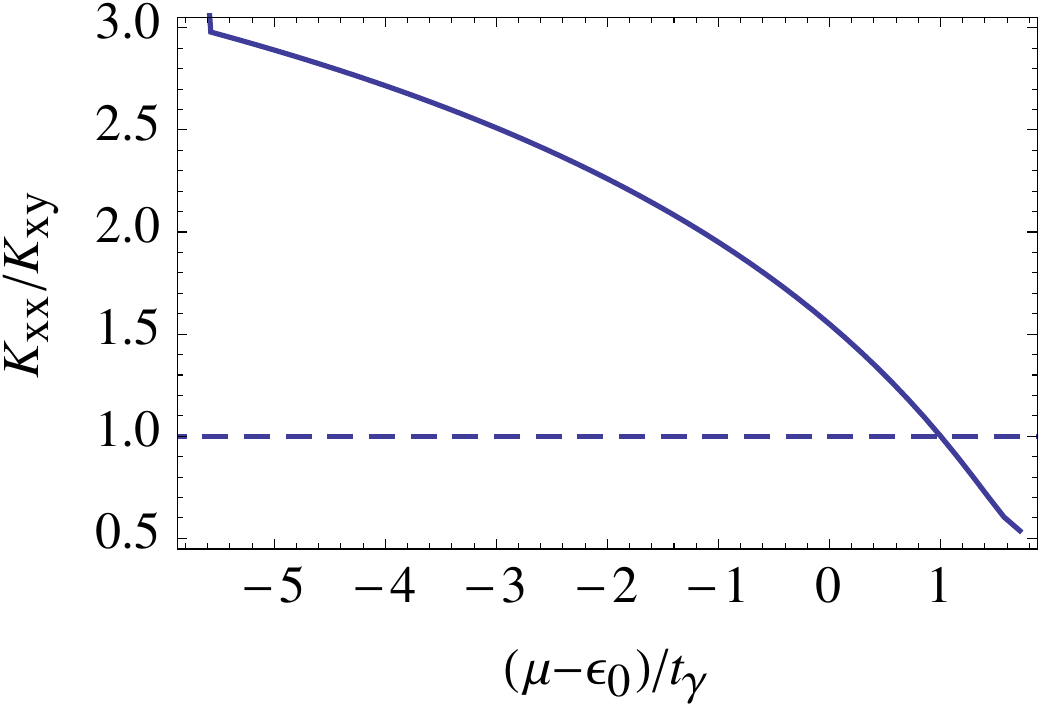} 
\caption{\label{Ks_ratio} Ratio of Ginzburg-Landau coefficients $K_{xx}/K_{xy}$ as a function of the filling $\mu-\epsilon_0$, computed in the weak-coupling limit for a tight-binding model. The remaining tight-binding parameter is chosen as $t'_{\gamma}/t_{\gamma} = 0.43 $. There is a threshold filling, $(\mu_c - \epsilon_0)/t_{\gamma} \approx  1$, above which $K_{xy}$ is bigger than $K_{xx}$. At $(\mu-\epsilon_0 )/t_{\gamma}= -5.71 $ the filling is zero (isotropic limit) with the ratio $K_{xx}/K_{xy}=3$. }
\end{figure}

We consider now the edge currents within the GL formulation. The order parameter at the edge can be characterized by 
the simplified boundary conditions (ignoring the vector potential) that the component $ \Delta_n = 0 $ and $ \partial_{n} \Delta_{\bar{n}} = 0 $ at the surface 
where $ n $ denotes the component perpendicular ($ \bar{n} $ parallel) and $ \partial_{n} $ is the derivative perpendicular to the surface. We restrict to the two main directions with $ \theta_{\rm Edge}= 0^{\circ} $ and $ 45^{\circ} $. An approximative spatial dependence of the order parameter is given by $ \Delta_n(r_n) = \Delta_0 \tanh(-r_n/\xi_n) $ and $ \Delta_{\bar{n}} = i \Delta_0 $ with $ r_n <0$ the coordinate perpendicular the surface located at $ r_n =0 $ and $ \xi_n $ the corresponding healing length. The current density $ j_n $ perpendicular to the surface vanishes naturally and the current density parallel is approximately given by
\begin{equation}
j_{\bar{n}} (r_n) \propto  K_n \Delta_0 \partial_n \left\vert \Delta_n(r_n) \right\vert = - \frac{K_n \Delta_0^2}{\xi_n} \cosh^{-2}(r_n/\xi_n)
\end{equation}
where $ K_n = \tilde{K}_{xx} $ for $ \theta_{\rm Edge} = 0^{\circ} $ and $ K_n = K_{xx} - K_{xy} $ for $ \theta_{\rm Edge} = 45^{\circ} $. Therefore, we always find a negative current density along a straight right edge (since $\partial_n \left\vert \Delta_n(r_n) \right\vert_{r_n=0} <0$ ). On the other hand, the sign of the current density parallel to a zigzag edge depends on the ratio $K_{xx}/K_{xy}$. Using the weak coupling expressions of the GL coefficients we plot this ratio on Fig. \ref{Ks_ratio} as a function of the filling $\mu-\epsilon_0 $ (the other tight-binding parameters are kept constant). We find a threshold filling $\mu_c$ above which the ratio $K_{xx}/K_{xy}$ becomes smaller than $1$ leading to a positive current density parallel to a zigzag edge (the other way around if the filling is below $\mu_c$). Therefore when $\mu<\mu_c$ the profile of the GL edge currents of an octagonal sample corresponds to Fig. \ref{EC_Low_High} (a), when $\mu>\mu_c$ it is given by Fig. \ref{EC_Low_High} (b). The full self-consistent solution of the quasi-classical and Ginzburg-Landau equations for the straight and zigzag edges will be presented in a future work. 


%
%

%



We briefly address here consequences of the filling dependence of the ratio $K_{xx}/K_{xy}$ for the structure of domain walls in the CPW state, parallel to one principal crystal axis, i.e. $\theta_{\mathrm{DW}}=0$. An approximate shape of domain wall can be given by $\Delta_{\nu}(x) = \alpha_{\nu} \Delta_0 \tanh \left(x/\tilde{\xi} \right)$ and $\Delta_{\bar{\nu}} = \beta_{\bar{\nu}} \Delta_0$ with $ \tilde{\xi} \propto \sqrt{K_{x \nu}} $, $\left(\alpha_{x},\beta_{y}\right) = \left(1,i\right)$ and $\left(\alpha_{y},\beta_{x}\right) = \left(i,1\right)$ for the two basic structures. Inserting this into the GL free energy we obtain, $ E_{DW} \propto \tilde{\xi} \propto  \sqrt{K_{x \nu}} $, 
%
%
such that the relative energy between the two types of domain wall is simply given by the ratio $K_{xx}/K_{xy}$. The stable domain wall structure (for $\theta_{\mathrm{DW}}=0$) is once more determined by the band filling. When $\mu<\mu_c$, the domain wall has the kink in the parallel component, here $\Delta_y$, and when $\mu>\mu_c$, the most stable domain wall has a kink in the perpendicular component, here $\Delta_x$. 
The discussion of stable domain wall shapes had so far been based on the low-filling properties of the CPW. Our extension here requires, therefore also a revision of the conclusions drawn based on domain wall structure, in particular, in the context of interference effects in Josephson contacts \cite{Maeno2006,Bahr,AnwarMaeno2013,BouhonSigrist2010}

%

Finally let us comment on a recent experiment aiming at detecting edge currents by a scanning magnetometer on the top-surface of  small cylinders (of radius $r\sim 5-10\mu$) of a highly pure single crystal of Sr$_2$RuO$_4$ \cite{Curran2014}. Since our analysis of a CPW state compatible with the $ \gamma$-band of Sr$_2$RuO$_4$ reveals the reversal of the current flow between straight and zigzag edges (Fig.~\ref{EC_Low_High}(b)), we expect that the edge currents may be strongly suppressed in circularly shaped samples of small radius. This might be responsible for the rather small upper bound for the currents, stated by Ref.~\cite{Curran2014}. Obviously also currents at extended edges might be influenced by our finding, in particular, if the edges are somewhat faceted \cite{Kirtley}. Note that a perfect square sample might be more suitable for a stronger conclusion. 



We conclude by noting that weak coupling functional renormalization group calculations suggest that next-to-nearest-neighbor pairing terms are very important in Sr$_2$RuO$_4$ \cite{Thomale2013}. Introducing them doesn't change the qualitative picture discussed in this work as long as the nearest-neighbor pairing coupling dominates. With growing next-to-nearest-neighbor pairing the CPW state exhibits a phase transition from the topological sector $C_1=\pm 1$ to $C_1=\mp3$ as a function of the filling. While the qualitative picture for the edge currents and our basic conclusions remain unchanged through such a transition, the bound state spectrum is strongly modified. A detailed discussion corresponding to this situation will be discussed elsewhere \cite{BouhonSigrist2014_Next}.







\begin{thebibliography}{}
\bibitem{Maeno-Mackenzie} A.P. Mackenzie and Y. Maeno, Rev. Mod. Phys. 75, 657 (2003).
\bibitem{Phys-Today} Y. Maeno, T.M. Rice. M. Sigrist, Physics Today 54, 42 (2001).
\bibitem{JPSJ-Maeno} Y. Maeno, S. Kittaka, T. Nomura, S. Yonezawa, and K. Ishida, J. Phys. Soc. Jpn. 81, 011009 (2012).
\bibitem{BalianWerthamer} R. Balian and N.R. Werthamer, Phys. Rev. {\bf 131}, 1553 (1963).
\bibitem{Leggett1975} A.J. Leggett, Rev. Mod. Phys. {\bf 47}, 331 (1975).
\bibitem{Balatskii1986} A.V. Balatskii, G.E. Volovik, and V.A. Konyshev, Zh. Eksp. Teor. Fiz. {\bf 90}, 2038-2056 (1986) [Sov. Phys. JETP {\bf 63}, 1194 (1986)]. 
\bibitem{Volovik1992} G. E. Volovik, \textit{Exotic properties of superfluid $^3$He}, Singapore: World Scientific, 1992.
\bibitem{Volovik1997} G. E. Volovik, JETP Lett. {\bf 66}, 522 (1997).
\bibitem{Volovik} G.E. Volovik, \textit{ The Universe in a Helium droplet}, Oxford University Press, 2003. 
\bibitem{MatSig2001} A. Furusaki, M. Matsumoto, and M. Sigrist, Phys. Rev. B {\bf 64}, 054514 (2001).
\bibitem{Kashiwaya2011} S. Kashiwaya, H. Kashiwaya, H. Kambara, T. Furuta, H. Yaguchi, Y. Tanaka, and Y. Maeno, PRL {\bf 107}, 077003 (2011). 
\bibitem{Kirtley} J. R. Kirtley, C. Kallin, C. W. Hicks, E.-A. Kim, Y. Liu, K. A. Moler, Y. Maeno, and K. D. Nelson: Phys. Rev. B \textbf{76}, 014526 (2007).
\bibitem{Budakian} J. Jang, D. G. Ferguson, V. Vakaryuk, R. Budakian, S. B. Chung, P. M. Goldbard, and Y. Maeno,
Science {\bf 331} 186 (2011).
\bibitem{Curran2014} P.J. Curran, S.J. Bending, W.M. Desoky, A.S. Gibbs, S.L. Lee, and A.P. Mackenzie, Phys. Rev. B {\bf 89}, 144504 (2014). 
\bibitem{Thomale2013}  Q. H. Wang, C. Platt, Y. Yang, C. Honerkamp, F.C. Zhang, W. Hanke, T.M. Rice and R. Thomale, EPL {\bf 104}, 17013 (2013).
\bibitem{Bergemann} C. Bergemann, A.P. Mackenzie, S.R. Julian, D. Forsythe, and D. Ohmichi, Adv. Phys. {\bf 52}, 639-725 (2003). 
\bibitem{Sigrist1991} M. Sigrist and K. Ueda, Rev. Mod. Phys. {\bf 63}, 239 (1991).
\bibitem{Maeno2006} F. Kidwingira, J.D. Strand, D.J. Van Harlingen and Y. Maeno, Science {\bf 314}, 1267 (2006).
\bibitem{Bahr} D. Bahr, PhD Dissertation, \textit{Probing the chiral domain structure and dynamics in Sr$_2$RuO$_4$ with nanoscale Josephson junctions}, University of Illinois, 2011.
\bibitem{AnwarMaeno2013} M.S. Anwar \textit{et al}, Scientific Reports {\bf 3}, 2480 (2013).
\bibitem{BouhonSigrist2010} A. Bouhon and M. Sigrist, N. J. Phys. {\bf 12}, 043031 (2010).
\bibitem{BouhonSigrist2014_Next} A. Bouhon and M. Sigrist, in preparation. 



\end{thebibliography}

\end{document}